\documentclass{article}

\usepackage{authblk}

\usepackage{amssymb,amsthm}
\usepackage{amsmath}
\usepackage{stmaryrd} 
\usepackage{setspace}
\usepackage[a4paper,top=25mm, bottom=25mm, left=25mm, right=25mm]{geometry}
\usepackage[small, margin=20pt]{caption}
\usepackage{enumitem}
\usepackage{color}
\usepackage[textsize=footnotesize,color=green!40]{todonotes}

\usepackage{eurosym}
\usepackage{fullpage}
\usepackage{palatino}
\usepackage{tikz}
\usepackage{verbatim}
\usepackage{fancybox}
\usepackage{xcolor}
\usepackage{subcaption}

\usepackage{algorithmic}
\usepackage{algorithm}
\floatname{algorithm}{Procedure}

\usepackage{soul}%strikethrough: st

\usepackage{hyperref}
\allowdisplaybreaks[4]

\newtheorem{claim}{Claim}

\theoremstyle{plain}
\newtheorem{theorem}{Theorem}%[section]
\newtheorem{lemma}[theorem]{Lemma}
\theoremstyle{definition}

\newtheorem{question}[theorem]{Question}
\newtheorem{observation}[theorem]{Observation}

\renewenvironment{proof}[1][Proof]{\begin{trivlist}
\item[\hskip\labelsep {\textit{#1}.}]}{\qed\end{trivlist}}

\newcommand{\cG}{\mathcal{G}}

\newcommand{\source}{{\sf s}}
\newcommand{\target}{{\sf t}}

\title{Recoloring graphs of treewidth $2$\thanks{This work was supported by ANR project GrR (ANR-18-CE40-0032).}}
\author[1]{Valentin Bartier}
\author[2]{Nicolas Bousquet}
\author[3]{Marc Heinrich}
\affil[1]{Univ. Grenoble Alpes, CNRS, Laboratoire G-SCOP, Grenoble-INP, Grenoble, France.\texttt{valentin.bartier@grenoble-inp.fr}}
\affil[2]{Universit\'e de Lyon, Universit\'e Lyon 1, LIRIS, UMR CNRS 5205, Lyon, France, France. E-mail: \texttt{nicolas.bousquet@univ-lyon1.fr}}
\affil[3]{School of Computing, University of Leeds, UK. \texttt{M.Heinrich@leeds.ac.uk}}

\date{}

\begin{document}

\maketitle

\begin{abstract}
Two (proper) colorings of a graph are adjacent if they differ on exactly one vertex. Jerrum proved that any $(d+2)$-coloring of any $d$-degenerate graph can be transformed into any other via a sequence of adjacent colorings. 
A result of Bonamy et al. ensures that a shortest transformation can have a quadratic length even for $d=1$. Bousquet and Perarnau proved that a linear transformation exists for between $(2d+2)$-colorings. It is open to determine if this bound can be reduced.

In this paper, we prove that it can be reduced for graphs of treewidth~$2$, which are $2$-degenerate. More formally, we prove that there always exists a linear transformation between any pair of $5$-colorings. This result is tight since there exist graphs of treewidth~$2$ and two $4$-colorings such that a shortest transformation between them is quadratic.
\end{abstract}

\section{Introduction}
Reconfiguration problems consist in finding step-by-step transformations between two feasible solutions of a problem such that all intermediate states are also feasible. Such problems model situations where a solution already in place has to be modified for a more desirable one while keeping some properties all along. Reconfiguration problems have been studied in various fields such as discrete geometry~\cite{BoseLPV18}, optimization~\cite{BBRM18} or statistical physics~\cite{mohar4}.
%In the last few years, reconfiguration problems of graph-related objects received a considerable attention such as independent sets~\cite{BonamyB17,LokshtanovM18} or matchings~\cite{BousquetHIM18,ItoKK0O17}. 
For a complete overview of the reconfiguration field, the reader is referred to the two recent surveys on the topic~\cite{Nishimura17,Heuvel13}. In this paper, our reference problem is graph coloring.

%Let $\Pi$ be a problem and $I$ be an instance of $\Pi$. The \emph{reconfiguration graph} is the graph where vertices are solutions of $I$ and where there is an edge between two vertices if one can transform the first solution into the other in one step (for graph colouring one step means modifying the colour of a single vertex\footnote{Note that recolouring operations have been studied, for instance recolouring using Kempe chains (see e.g.~\cite{BonamyBFJ19}). In this article, we focus on single vertex recolourings.}).
%Given a reconfiguration problem, several questions may arise. (i) Is it possible to transform any solution into any other, i.e. is the reconfiguration graph connected? (ii) If yes, how many steps are needed to perform this transformation, i.e. what is the diameter of the reconfiguration graph? 
In this work, we will focus on the diameter of the reconfiguration graph. The diameter of the reconfiguration graph plays an important role, for instance in random sampling, since it provides a lower bound on the mixing time of the underlying Markov chain (and the connectivity of the reconfiguration graph ensures the ergodicity of the Markov chain\footnote{Actually, it only gives the irreducibility of the chain. To get the ergodicity, we also need the chain to be aperiodic. For the chains associated to proper graph colorings, this property is usually straightforward.}). Since proper colorings correspond to states of the anti-ferromagnetic Potts model at zero temperature, Markov chains related to graph colorings received a considerable attention in statistical physics and many questions related to the ergodicity or the mixing time of these chains remain widely open (see e.g.~\cite{ChenDMPP19,frieze2007survey}).

\paragraph{Graph recoloring.}
All along the paper $G=(V,E)$ denotes a graph, $n:=|V|$ and $k$ is an integer. For standard definitions and notations on graphs, we refer the reader to~\cite{Diestel}.
A \emph{(proper) $k$-coloring} of $G$ is a function $\sigma : V(G) \rightarrow \{ 1,\ldots,k \}$ such that, for every edge $xy\in E$, we have $\sigma(x)\neq \sigma(y)$. Throughout the paper we only consider proper colorings and will then omit the proper for brevity. The \emph{chromatic number} $\chi(G)$ of a graph $G$ is the smallest $k$ such that $G$ admits a $k$-coloring.
Two $k$-colorings are \emph{adjacent} if they differ on exactly one vertex. The \emph{$k$-reconfiguration graph of $G$}, denoted by $\cG(G,k)$ and defined for any $k\geq \chi(G)$, is the graph whose vertices are $k$-colorings of $G$, with the adjacency relation defined above. %Note that two colourings equivalent up to colour permutation are distinct vertices in the reconfiguration graph.
Cereceda et al. provided an algorithm to decide whether, given two $3$-colorings, one can be transformed into the other in polynomial time, and characterized graphs for which $\cG(G,3)$ is connected~\cite{Cereceda09,CerecedaHJ11}.
Given any two $k$-colorings of $G$, it is $\mathbf{PSPACE}$-complete to decide whether one can be transformed into the other for $k \geq 4$~\cite{BonsmaC07}. 

The \emph{$k$-recoloring diameter} of a graph $G$ is the diameter of $\cG(G,k)$ if $\cG(G,k)$ is connected and is equal to $+\infty$ otherwise. In other words, it is the minimum~$D$ for which any $k$-coloring can be transformed into any other one through a sequence of at most~$D$ adjacent $k$-colorings.
Bonsma and Cereceda~\cite{BonsmaC07} proved that there exists a family $\mathcal{G}$ of graphs and an integer $k$ such that, for every graph $G \in \mathcal{G}$ there exist two $k$-colorings whose distance in the $k$-reconfiguration graph is finite and super-polynomial in $n$.

Cereceda conjectured that the situation is different for degenerate graphs.
A graph $G$ is \emph{$d$-degenerate} if any subgraph of $G$ admits a vertex of degree at most~$d$. In other words, there exists an ordering $v_1,\ldots,v_n$ of the vertices such that for every $i \le n$, the vertex $v_i$ has at most $d$ neighbors in $v_{i+1},\ldots,v_n$.
It was shown independently by Dyer et al.~\cite{dyer2006randomly} and Cereceda et al.~\cite{Cereceda09} that for any $d$-degenerate graph $G$ and every $k \geq d+2$, $\cG(G,k)$ is connected. However the (upper) bound on the $k$-recoloring diameter given by these constructive proofs is of order $c^{n}$ (where $c$ is a constant). 
%Note that the bound on $k$ cannot be decreased and maintain the existence of a transformation since $\cG(K_n,n)$ is not connected (and $K_n$ is $(n-1)$-degenerate). 
Cereceda~\cite{Cereceda} conjectured that the the diameter of $\cG(G,k)$ is of order $\mathcal{O}(n^2)$ as long as $k \ge d+2$. If correct, the quadratic function is sharp, even for paths or chordal graphs as proved in~\cite{BonamyJ12}. Bousquet and Heinrich~\cite{BousquetH19} proved that the diameter of $\cG(G,k)$ is $n^{d+1}$, improving the $2^n$ upper bound of~\cite{dyer2006randomly}. The conjecture is known to be true for a few graph classes such as chordal graphs~\cite{BonamyJ12} or bounded treewidth graphs~\cite{BonamyB13,Feghali19}. 

%In this work, we want to determine when the diameter of the reconfiguration graph becomes linear. 
One can wonder what happens if $k$ increases. No non trivial lower bound on the diameter of $\cG(G,k)$ is known for $d$-degenerate graphs $G$ when $k \ge d+3$. Bousquet and Perarnau~\cite{BousquetP16} showed that the diameter of $\cG(G,k)$ is linear when $k \ge 2d+2$. Bartier and Bousquet~\cite{BousquetB19} proved that the diameter of $\cG(G,k)$ is linear for bounded degree chordal graphs. The problem remains open for bounded treewidth graphs or general chordal graphs.

The class of planar graphs also received a considerable attention. Feghali proved that $\cG(G,k)$ is almost linear (precisely $O(n \cdot Polylog(n))$) when $G$ is a planar graph and $k \ge 8$~\cite{Feghali19+}. Dvorak and Feghali also showed that it becomes linear when $k \ge 10$~\cite{DvorakF20+}. One can naturally ask for which graph classes the diameter of $\cG(G,k)$ becomes linear when $k \ge d+2$. 

In this paper, we investigate the class of outerplanar graphs and more generally the graphs of treewidth $2$. Since graphs of treewidth $2$ are $2$-degenerate, the result of~\cite{BousquetP16} ensures that there exist linear transformations between any pair of $6$-colorings. We also know that such a result is impossible for $4$-colorings~\cite{BonamyJ12} (the diameter is quadratic). So the only open case is when $k=5$. In this note, we answer this question and show:

\begin{theorem}\label{thm:main}
Let $G$ be graph of treewidth at most $2$ and $k=5$. There exists a constant $c$ such that, for every pair of $5$-colorings $\alpha,\beta$ of $G$, there exists a transformation from $\alpha$ to $\beta$ recoloring each vertex at most $c$ times.
\end{theorem}

Note that since outerplanar graphs have treewidth $2$ and the quadratic diameter for $k=4$ obtained in~\cite{BonamyJ12} also holds for outerplanar graphs, it also completely characterizes the recoloring diameter of outerplanar graphs.

One can naturally ask if the results of Theorem~\ref{thm:main} can be extended further. In particular we ask the two following questions, where the second generalizes the first one:

\begin{question}
Does there exists a constant $C$ such that, for every graph $G$ of treewidth at most $d$, the diameter of $\cG(G,k)$ is linear when $k \ge d +C$? 
\end{question}

\begin{question}
Does there exists a constant $C$ such that, for every $d$-degenerate graph $G$, the diameter of $\cG(G,k)$ is linear when $k \ge d +C$? 
\end{question}

\paragraph{Outline of the paper}
Before going into the details of the proof, let us describe from a high level point of view how the proof works, and how the paper is organized. The first step in the proof of Theorem~\ref{thm:main} is to show that it is sufficient to prove the result for chordal graphs with cliques of size at most~$3$, instead of graphs of treewidth~$2$. This reduction to chordal graphs is done in Section~\ref{sec:redChordal}. Proving that the result holds for chordal graphs is the main technical part of the paper, and is done in Section~\ref{sec:bestChoice}. The proof for chordal graphs is algorithmic in nature. We first describe the procedure, introduced in~\cite{BousquetP16}, which builds a recoloring sequence by making greedy choices. We prove that the recoloring sequences produced by this procedure recolor each vertex of the graph a constant number of times. This is done by first proving some properties of the sequences produced by this algorithm in Section~\ref{sec:prop}, and then showing in Section~\ref{sec:mainlemma} how these properties can be used to bound the number of times that each vertex is recolored. The core of the proof proceeds by contradiction. Assuming that a vertex is recolored a large number of times puts some very strong constraints on the recoloring sequence on its neighbor. This in turn is used to show that the algorithm would recolor $x$ a much smaller number of times in this sequence.

\section{Reduction to chordal graphs}
\label{sec:redChordal}

Let $G=(V,E)$ be a graph of treewidth at most $d$ and $T$ be a tree decomposition of $G$. A tree decomposition $(T,\mathcal{X})$ is a tree $T$ given with a bag function $\mathcal{X}$ such that, for every $u \in V(T)$, $\mathcal{X}(u)$ is a subset of $V(G)$, for every edge $e=xy$ in $E(G)$, there exists a node $u$ of $T$  such that $\{ x, y\} \subseteq \mathcal{X}(u)$. Moreover, for every $x \in V(G)$, the subset of nodes containing $x$ in their bags is connected.

The \emph{clique number of $G$}, denoted by $\omega(G)$ is the maximum size of a clique in $G$. The following lemma is sufficient to prove Theorem~\ref{thm:main}. 
%Lemma~\ref{lem:mainchordal} essentially re-states Theorem~\ref{thm:main} with weaker assumptions and conclusions.

\begin{lemma}\label{lem:mainchordal}
There exists a constant $c$ such that for every chordal graph $H$ with $\omega=3$, there exists a transformation from any $5$-coloring of $H$ into a $3$-coloring of $H$ by recoloring each vertex at most $c$~times. 
\end{lemma}

The proof of Lemma~\ref{lem:mainchordal} is postponed to Section~\ref{sec:mainlemma}. Let us explain how we can use it to prove Theorem~\ref{thm:main}

\begin{proof}[Proof of Theorem~\ref{thm:main}]
%We actually want to show that we can recolor $\alpha$ into a $3$-coloring $\beta$ of $G$ in a linear number of steps using Algorithm~\ref{TODO}. 
Let $G$ be a graph of treewidth at most $2$ and $\alpha, \beta$ be two $5$-colorings of $G$. Let $T$ be a tree decomposition $T$ of $G$ that can be found in linear time by~\cite{Bodlaender93}.

Let us first transform $G$ into a chordal graph $H$ with clique number at most three.
The transformation is based on a trick used in~\cite{Feghali19}. For every bag $B$ of $T$, we merge the vertices of~$B$ colored the same in $\alpha$. We repeat until we obtain a graph $H'$ such that no bag contains two vertices of the same color. Note that $tw(H') \leq tw(G)$ since we only identify vertices belonging to the same bags. Since we merged vertices colored the same in $\alpha$, the coloring $\alpha'$ which is the coloring of $H'$ where a vertex receives the color of its color class in $G$ is proper and well-defined.
Note that, in $H'$, vertices belonging to the same bag receive distinct colors. So $\alpha'$ is also a coloring of $H$ which is the chordal graph whose clique tree is $T_{H'}$. In other words, $H$ is the graph obtained from $H'$ by transforming every bag of the tree decomposition of $H'$ into cliques.
%We then add edges between any pair of non-adjacent vertices that are contained in the same bag. Let $H'$ be the graph obtained after this last procedure and $\alpha'$ be the coloring induced by $\alpha$ on $H$. Note that $H'$ is a chordal graphs. 

By Lemma~\ref{lem:mainchordal}, $\alpha'$ can be transformed into a $3$-coloring of $H$ by recoloring every vertex at most $c$ times.
Feghali observed in~\cite{Feghali19} that it implies that, in $G$, we can transform $\alpha$ into a $3$-coloring $\gamma_1$ of $G$ by recoloring every vertex of $G$ at most $c$ times.
With a similar argument, we can transform $\beta$ in a $3$-coloring $\gamma_2$ by recoloring every vertex at most $c$ times. The following claim concludes the proof of Theorem~\ref{thm:main} by taking $d=2$.

\begin{claim}\label{lem:freecolors}
Let $G$ be a graph of treewidth at most $d$ and $\gamma_\source,\gamma_\target$ be two $(d+1)$-colorings of $G$ using colors $\{1,\ldots,d+1 \}$. If $k \ge 2d+1$, $\gamma_\source$ can be transformed into $\gamma_\target$ by recoloring every vertex at most twice. 
\end{claim}
We perform the following recolorings. We denote by $X_i$ the subset of vertices colored with~$i$ in $\gamma_\source$. For every $i \le d$, recolor one by one all the vertices of $X_i$ with color $d+1+i$. Then recolor the vertices of $X_{d+1}$ with their target colors. Finally recolor all the vertices of $X_i$ for $i \le d$ with their target color. The resulting coloring is $\gamma_\target$ and we claim that at every step, the current coloring is proper.
Indeed, during the first phase, the color classes are the same (we simply change the color classes of the $X_i$s). When $X_{d+1}$ is recolored, all the vertices that receive a color in $\{ 1,\ldots,d+1 \}$ in the current coloring have their final color. The property is kept all along the rest of the recoloring algorithm, which completes the proof since $\gamma_\target$ is proper. $\Diamond$
\end{proof}

So, in what follows, our goal is to prove Lemma~\ref{lem:mainchordal}.

\section{Best choice algorithm}\label{sec:bestChoice}

Our proof of Lemma~\ref{lem:mainchordal} is algorithmic and is based on an algorithm already used for instance in~\cite{BousquetP16}. Our main contribution consists in making a careful analysis of its behavior for chordal graphs with clique number at most~$3$, and showing that the recoloring sequence produced by the algorithm recolors each vertex a constant number of times. We start by describing the algorithm and its properties. 

In the following, given a graph $G$ and two colorings $\alpha$ and $\beta$ of $G$, a recoloring sequence $\mathcal S$ is a sequence $s_1s_2\ldots s_t$ where each $s_i$ is a pair $(v, c)$ describing the vertex $v$ and the color $c$ which characterize the $i$-th step of the sequence. All the intermediate colorings described by this sequence must be proper. The \emph{restriction of $\mathcal{S}$ to $X$}, denoted $\mathcal S_{|X}$, is the recoloring sequence obtained from $\mathcal S$ by keeping only recolorings of vertices in $X$. In particular if $X$ is a single vertex $v$, then $\mathcal S_{|v}$ denotes the sequence of colors taken by the vertex $v$ in the recoloring sequence $\mathcal S$. With this notation, $|S_{|v}|$ is the number of times the vertex $v$ is recolored in the sequence $\mathcal S$. 

Part of our proof relies on the identification of patterns in a recoloring sequence $\mathcal S$. Given a sequence of vertices $v_1\ldots v_\ell$, we say that $\mathcal S$ contains this pattern if there is an index $i$ such that for all $j \leq \ell$, $s_{i +j}$ recolors the vertex $v_j$. Given an integer $r$, we also denote by $v_i^{\geq r}$ the pattern where $v_i$ is recolored $r$ times in a row (without any other vertex interleaved in $\mathcal S$), and $(v_1\ldots v_\ell)^r $ to denote that the pattern $v_1 \ldots v_\ell$ is repeated $r$ times in a row in the sequence $\mathcal S$.

Consider a $d$-degenerate graph $G$, two $k$-colorings $\alpha, \beta$ of $G$ with $k \geq d+2$, and  a vertex $u$ of $G$ with degree at most $d$.  Let $\alpha'$ and $\beta'$ be the restrictions of $\alpha$ and $\beta$ to $G - u$ and $\mathcal{S}$ a recoloring sequence from $\alpha'$ to $\beta'$. Let us now explain how we can extend this recoloring sequence to the whole graph.

Let $t_1, \ldots, t_{\ell}$ be all the steps where the color of any neighbor of $u$ is modified in $\mathcal{S}$, and let $c_{t_i}$ be the new color assigned to the recolored neighbor at step $t_i$. A \emph{best choice} for $u$ at step $t \in \{ t_1, \ldots, t_\ell\}$ is:
\begin{itemize}
    \item the color $\beta(u)$ if $\beta(u)$ is distinct from $c_{t},\ldots,c_{t_\ell}$;
    \item or any valid choice for $u$ at step $t$ which is distinct from  $\{ c_{t_i}$ with $t_i \ge t \}$ otherwise;
    \item or the valid choice for $u$ at step $t$ that appears the latest in the sequence $(c_{t_i})_{t_i \ge t}$ otherwise.
\end{itemize}

The Local Best Choice for $u$ extends the sequence $\mathcal{S}$ by recoloring only the vertex $u$. When a neighbor $v$ of $u$ is recolored in $\mathcal{S}$ with the current color of $u$, we add a recoloring for $u$ just before this recoloring and we recolor $u$ with a best choice. We do not perform any other recoloring for $u$ except at the very last step to give it color $\beta(u)$ if needed. 

Let $G$ be a chordal graph and $v_1,\ldots,v_n$ be a degeneracy ordering of $G$.
The Best Choice Recoloring Algorithm is the algorithm which consists in making the Local Best Choice successively on $v_n,\ldots,v_1$. We call a \emph{best choice recoloring} a sequence $\mathcal S$ which can be obtained from this procedure for some degeneracy ordering of $G$. Observe that by construction, if $\mathcal S$ is a best choice recoloring for some graph $G$ and some degeneracy ordering $v_1, \ldots, v_n$, then for every index $i \geq 0$, if $V_i = \{v_i, \ldots, v_n \}$, then $\mathcal S_{|V_i}$ is a best choice recoloring for $G[V_i]$.

%Before proving our main results let us start with some basic properties on recolorings of degenerate graphs.

\subsection{Properties of the Best Choice Algorithm}
\label{sec:prop}

In what follows, we always assume that $G$ is a chordal graph with $\omega=3$ and $k=5$. The ordering $v_1, \ldots , v_n$ is a perfect elimination ordering of $G$, and $\mathcal S = s_1 \ldots s_t$ is a best choice recoloring for this ordering between two $5$-colorings $\alpha$ and $\beta$ of $G$. We denote by $V_i$ the set $\{v_i, \ldots, v_n\}$, and $G_i = G[V_i]$. The perfect elimination ordering of $G$ can be used to construct an acyclic orientation of the graph, with the edge $v_i v_j$ oriented towards $v_j$ if $j > i$. With this convention, each vertex of the graph has at most~$2$ out-neighbors, and for every vertex $v$ of $G$, we denote by $N^+(v)$ these two out-neighbors, and $N^+[v] = N^+(v) \cup \{v\}$ the inclusive out-neighborhood. Because the ordering is a perfect elimination ordering, the out-neighbors of a vertex are adjacent.

Given a vertex $v$, the step $i$ of the sequence $\mathcal S$ is \emph{saved for} $v$ if $s_i$ recolors a vertex $w \in N^+(v)$ and one of the following holds:
\begin{itemize}
	\item $v$ is not recolored at steps $1, \ldots, i$;
	\item $v$ is not recolored at steps $i, \ldots, t $; 
	\item the two steps preceding $s_i$ in $\mathcal S_{|N^+[v]}$ do not recolor $v$.
\end{itemize}

Informally, in the "worst case" scenario, when we apply the Best Choice Algorithm to extend a recoloring sequence to a new vertex $v$, whenever we have to recolor $v$ because of one of its neighbors there are two possible choice of color to recolor $v$ with. By choosing the correct color, the algorithm recolors $v$ once every two steps at most. This is formalized in the following observation:

\begin{observation}
	\label{obs:half-time}
	For every vertex $v$ and $w \in N^+(v)$, the patterns $vv$ never occur in $\mathcal S_{|N^+[v]}$, and the pattern $v w v$ can occur at most once, at the very end of the sequence $\mathcal S_{|N^+[v]}$.
\end{observation}

Saved steps quantify how many additional steps are saved by the algorithm compared to the worst case scenario. More precisely, we will show that every two saved recolorings for $x$ reduce the number of times $x$ is recolored by one compared with the worst case scenario. In order to prove this result, we will need one additional notion. We say that a step $i$ recoloring a vertex $v$ is \emph{caused by} a vertex $w$ if the recoloring step following $s_i$ in $\mathcal S_{N^+(v)}$ recolors $w$. Note that by construction, this recoloring step recolors $w$ with the color that $v$ had just before the transformation $s_i$. Moreover, all the steps recoloring $v$, except possibly the last one where $v$ is given its target color, are caused by one of the out-neighbor of $v$.

Let us start by proving a simple inequality on the number of times that a vertex $v$ is recolored.

\begin{lemma}
	\label{obs:nb-recol}
	For every vertex $v$ of, if $r$ is the number of saved recolorings for $v$, then the following inequality holds:
	 	$$|S_{|v}| \leq 1 - \frac r 2 + \left\lceil \frac{\sum_{w \in N^+(v)} |\mathcal S_{|w}| }{2} \right\rceil \;.$$
\end{lemma}

\begin{proof}
The statement is proved by induction on $m:= \sum_{w \in N^+(v)} |S_{|w}|$. If $v$ is recolored at most once during the sequence (which is the case when $m=0$, i.e., when its out-neighbors are not recolored) the conclusion follows immediately. Hence, let us assume that $v$ is recolored at least twice. If $v$ is not the first recolored vertex of $\mathcal{S}_{|N[v]}$, then the first recoloring is saved. If we consider the coloring $\alpha'$ obtained after this first recoloring, and $\mathcal S'$ the subsequence of $\mathcal S$ which recolors $\alpha'$ to $\beta$, then by induction, $v$ is recolored at most $\frac{(m-1)-(r-1)}{2}+1$ times in $\mathcal S'$, and consequently also in $\mathcal S$. In this case the induction step holds, consequently we can assume in the rest of the proof that $v$ is the first vertex recolored in $\mathcal S_{|N^+[v]}$.

Let us write $\mathcal S_{|N^+[v]} = s'_1 \ldots s'_\ell$. We know that $s'_1$ recolors $v$, let us write $c_0$ and $c_1$ the colors of $v$ respectively before and after the transformation $s'_1$. Since $v$ is recolored at least twice, this recoloring must be caused by one of its out-neighbors say $w$, and $w$ must be recolored in $s'_2$ with the color $c_0$. Consequently, there are still two colors that did not appeared in $N^+[v]$ in the transformation up to step $2$. By Observation~\ref{obs:half-time}, either $s'_3$ recolors $v$, and in this case is the last step of $\mathcal S_{|N^+[v]}$ (i.e., $\ell = 3$); or $s'_3$ does not recolor $v$. In the first case we have $\sum_{w \in N^+(v)} |S_{|w}| = 1$, and since $v$ is recolored twice, the inequality holds. In the second case we consider the coloring $\alpha'$ obtained after the transformation $s'_3$, and let $\mathcal S'$ be the subsequence starting after the step $s'_3$ recoloring $\alpha'$ into $\beta$. We know that $v$ is recolored at one less times in $\mathcal S'$ than in $\mathcal S$. Moreover, by definition neither $s'_2$ nor $s'_3$ are saved recolorings for $v$, hence the number of saved recolorings for $v$ in $\mathcal S'$ is still equal to $r$. Using the induction hypothesis on $\mathcal S'$, we know that $v$ is recolored at most $ \frac{m - 2 - r}{2} +1$ in $\mathcal S'$, and since $v$ is recolored one additional time in $\mathcal S$, the induction step holds.
\end{proof}

Hence in order to prove that a vertex is not recolored to many times, it is sufficient to show that there exists a sufficient number of saved recolorings. The following lemma gives a sufficient condition for a saved recoloring to appear.

\begin{lemma}\label{obs:skip-recol}
Let $u$ be a vertex of $G$ and $v, w \in N^+(u)$. We write $\mathcal S_{|N^+[u]} = s'_1, \ldots s'_\ell$. Assume that the step $i$ of $ S_{|N^+[u]}$ recolors $u$ and is caused by $v$, and step $i+1$ is caused by $w$. Then either $i+3 \geq \ell$  or the step $i+3$ of $S_{|N^+[u]}$  is saved for $u$.
\end{lemma}
\begin{proof}
Let $c_1$ (resp. $c_2$, resp. $c_3$) denotes the color of $u$ (resp. $v$, resp. $w$)  just before step $i$. By definition, at step $i+1$ the vertex $u$ takes the color $c_1$ (since the step $i$ is caused by $u$) and at step $i+2$, $w$ takes the color $c_2$ (since $i+1$ is caused by $w$). Thus, after step $i+2$, the only colors which appeared for $v$ and $w$ at steps $i,i+1$ and $i+2$ are the colors $c_1,c_2,c_3$. Since we make the best choice for $u$ and there remain two valid colors, $u$ is not recolored at step $i+3$ (and then is saved for $u$) except if it is the last step of $\mathcal S_{|N^+[u]}$ where $v_1$ receives its final color.
\end{proof}

\begin{lemma}\label{obs:h-cstr}
Let $u$ be a vertex of $G$, and $v, w \in N^+(v)$. If $\mathcal{S}_{|N[u]}$ contains the pattern $u vw^{>0}u$ then the initial, intermediate and final colors of $u$ in the subsequence corresponding to this pattern are pairwise distinct.
\end{lemma}
\begin{proof}
If the three colors are not pairwise distinct then only the initial and final colors of $u$ can be the same. Let $c_1$ be the initial color of $u$. Let $s'_1 \ldots s'_\ell$ be the subsequence of $\mathcal{S}_{|N[v]}$ corresponding to the pattern, then $s'_1$ is caused by $v$, which implies that $v$ is recolored in $s'_2$ with $c_1$. Since $v$ and $w$ are adjacent, all the subsequent recolorings of $w$ and $u$ use a color different from $c_1$, and in particular, the final color of $v$ is different from $c_1$.
\end{proof}

\begin{lemma}\label{lem:badtriples}
Let $x, u, v, w$ be vertices of $G$ such that $N^+(x) = \{u,v\}$ and $N^+(u) = \{v, w\}$. If $i$ is the first index such that $x = v_i$, let us write $c = \max_{i' > i} |\mathcal S_{|v_{i'}}|$. If $x$ is recolored at least $c - 1$ times in $\mathcal{S}$, then:
\begin{enumerate}
\item the pattern $u w v u$ appears at least $c-34$ times in $\mathcal S_{|N^+(u)}$;
\item in the sequence of colors of $x$, there are at most $74$ indices where three consecutive colors are not pairwise distinct.
\end{enumerate}
\end{lemma}
\begin{proof}
Let us prove the two points of the lemma successively. 
\begin{enumerate}
	\item Let us write $\mathcal S_{|N^+(u)} = s'_1, \ldots, s'_\ell$. Let us start by showing that few recolorings of $u$ are caused by $v$. Let $i$ be an index such that the recoloring $s'_i$ of $u$ is caused by the recoloring $s'_{i+1}$ of $v$. 
	
	First observe that $x$ is not recolored between $s'_i$ and $s'_{i+1}$. Indeed, let us denote by $c$ the color of $u$ before $s'_i$. Since the step $i$ is caused by $v$, this means that $v$ is recolored with $c$ in $s'_{i+1}$. Moreover, since $x$ and $u$ are adjacent, $x$ is not colored $c$ before $s'_i$. This implies that $x$ is not recolored between $s'_i$ and $s'_{i+1}$, as it would be a recoloring caused by $v$, which can only happen if $v$ is recolored with the color of $x$ which is not the case here.
	
	If $s'_i$ is the first recoloring of a vertex in $N^+[x]$, then it is saved for $x$. Otherwise, the recoloring preceding $s'_i$ in $\mathcal S_{|N+[x]}$ can be either:
	\begin{itemize}
		\item a recoloring of $x$, in which case it is caused by $u$, and by Lemma~\ref{obs:skip-recol}, this implies that either $s'_{i+1}$ is the last, or before last, step of $\mathcal S_{|N^+[x]}$, or the step following $s'_{i+1}$ in $\mathcal S_{|N^+[x]}$ is saved for $x$.
		\item a recoloring of either $u$ or $v$, and in this case $s'_{i+1}$ is saved for $x$.
	\end{itemize}
	This implies that every time (except once) there is a recoloring of $u$ caused by $v$, we can find a saved step for $x$, and all these saved steps are different. Moreover, since $x$ is recolored at least $c -1$ times, by Lemma~\ref{obs:nb-recol} there are at most $5$ saved recolorings for $x$, and consequently by the observation above there are at most $6$ recolorings of $u$ caused by $v$.
	
	Since $x$ is recolored at least $c-1$ times, we know by Lemma~\ref{obs:nb-recol} that $u$ is recolored at least $c - 3$ times. At most $6$ of these recolorings are caused by $v$ by the argument above. Hence, except the last one, the other $c - 10$ recolorings of $u$ are caused by $w$. Moreover, since $w$ is recolored at most $c$ times, it also means that there are at most $10$ recolorings of $w$ which do not cause a recoloring of $u$. 
	
	Let us consider $i$ and $j$ such that $s'_i$ and $s'_j$ are two consecutive recolorings of $u$ in $\mathcal S_{N^+[u]}$. Our goal is to show that except for a small number of choices of $i$, we have $j = i +3$, and $s'_i \ldots s'_j$ matches the pattern $uwvu$. Observe that the following properties hold:
	\begin{itemize}
		\item By Observation~\ref{obs:half-time}, we have $j \geq i +2$, and $j = i+2$ can only happen once at the end of the sequence.
		\item If $j > i+3$, i.e., if there are at least three recolorings of $v$ or $w$ between $s_i$ and $s_j$, then at least one of these recolorings is saved for $u$. This can happen at most $9$ times since $u$ is recolored at least $c-3$ times, and using Lemma~\ref{obs:nb-recol}.
		\item If $j = i +3$ and $s'_{i+1}$ does not recolor $w$, then we have a recoloring of $u$ which is not caused by $w$, which can happen at most $10$ times by the arguments from the previous paragraph.
		\item Finally, if $j = i+3$ and $s'_{i+1}$ recolors $w$, and if $s'_{i+2}$ does not recolor $v$, then it must recolor $w$ (it cannot recolor $u$ by the assumption that $j = i+3$), in which case we have a recoloring of $w$ which does not cause a recoloring of $u$, which can happen at most $10$ times.
	\end{itemize}
	
	Combining all the points above, and since $u$ is recolored at least $c-3$ times, the pattern $u w v u$ occurs at least $c - 4 - 1 - 9 - 10 - 10 = c - 34$ times, proving the first point of the lemma.

	\item Let us now consider the second point. Since $x$ is recolored at least $c-1$ times while its at most two out-neighbors are recolored at most $c$ times, then by Lemma~\ref{obs:nb-recol}, there are at most $5$ saved recolorings for $x$. Let us write $\mathcal S_{|N^+[x]} = s'_1\ldots s'_\ell$, and consider two indices $i < j$ such that $s'_i$ and $s'_j$ are two consecutive recolorings of $x$. Again, by Observation~\ref{obs:half-time}, we have $j \geq i+2$, and $j = i+2$ can occur only once at the end of the sequence. If $j > i+3$, then $s_{j - 1}$ is saved for $x$, which can happen at most $5$ times. In all the other cases, there are exactly two recolorings of either $u$ or $v$ between $s_i$ and $s_j$. By the point 1. above, we know that the pattern $u w v u$ appears at least $c - 34$ times in $\mathcal S_{| N^+[u]}$. This implies that in the sequence $\mathcal S_{|\{u, v\}}$ there are at most $2 \times 34$ times where either two consecutive $u$ or two consecutive $v$ appear as a pattern. Hence, in all but at most $68 +5+1 = 74$ occurrences, the subsequence $s_i \ldots s_j$ matches one of the patterns $x u v x$ or $x v u x$. By Lemma~\ref{obs:h-cstr} the three colors taken by $x$ during this portion of the recoloring sequence are all different, which proves the second point of the lemma.
\end{enumerate}
\end{proof}

\subsection{Proof of Lemma~\ref{lem:mainchordal}}
\label{sec:mainlemma}

%\begin{lemma}
%Let $c>TODO$.
%Let $G$ be a chordal graph and $x$ be a simplicial vertex. Assume moreover that for $G[V \setminus x]$, every vertex is recolored at most $c$ times using Algorithm~\ref{algo:TODO}.

%Then for $G[V]$, every vertex is recolored at most $c$ times using Algorithm~\ref{algo:TODO}.
%\end{lemma}
Let $c$ be a constant equal to $542$. In order to prove Lemma~\ref{lem:mainchordal}, we will show that a best choice recoloring sequence recolors each vertex at most $c$ times. This is proved by induction on the number of vertices of $G$. This is clearly true when $G$ contains a single vertex since the sequence will recolor this vertex at most once. Assume that the conclusion holds for all the chordal graphs with clique number at most three on $n$ vertices. Let $G$ be a graph with $n+1$ vertices and let $x$ be the first vertex in the elimination ordering.  Assume by contradiction that $x$ is recolored $c+1$ times in a best choice recoloring sequence for this elimination ordering. Using the induction hypothesis, all the vertices but $x$ are colored at most $c$ times.
 
If $x$ has a single neighbor, then by Lemma~\ref{obs:nb-recol}, it is recolored at most $\frac c 2 +1 \leq c$ times, a contradiction. Hence, we can assume that $x$ has two neighbors $y$ and $z$. Again, using Lemma~\ref{obs:nb-recol}, both $y$ and $z$ are recolored at least $c-1$ times in $\mathcal{S}$. Since $G$ is chordal, $yz$ is an edge of $G$, and we can assume without loss of generality that $z$ is an out-neighbor of $y$. Let $y_1$ be the second out-neighbor of $y$ (if it exists).

By the first point of Lemma~\ref{lem:badtriples}, there are at most $34$ recolorings of $y$ in $\mathcal{S}_{|N^+[y]}$ where the subsequence starting at this point does not match the pattern $yy_1zy$. Moreover, by the second point of Lemma~\ref{lem:badtriples} applied to $z$, there are at most $74$ triplets of consecutive colors of $z$ which are not composed of pairwise distinct colors. Since both $y$ and $z$ are recolored at least $c-1$ times, there exists a subsequence $\mathcal{S}'$ of $\mathcal{S}_{N^+[y]}$ of the form $(yy_1z)^{c'}$ where $c':= \lfloor \frac{c-2}{108} \rfloor = 5$ and where three consecutive colors of $z$ are always pairwise distinct.

For $X\in \{x, y, z, y_1\}$, let us denote by $c^X_i$ the color of $X$ after the $i$-th recoloring of $X$ in the sequence $\mathcal S'$, with the convention that $c^X_0$ is the initial color of $X$.

Note that all the recolorings of $y$ are caused by $y_1$ in this subsequence, consequently, $c^{y_1}_{i+1} = c^y_i$ for all $i \leq c'$. Since we choose the best color for $y$ when we recolor $y$ from $c^y_{i+1}$ to $c^y_{i+2}$, the set of colors $c^{y_1}_{i+1},c^{y_1}_{i+2},c^z_{i+1},c^z_{i+2}$ are pairwise distinct. Indeed,  if they were not, the algorithm could have chosen an other color for $y$, which would postponed the recoloring of $y$, a contradiction with the fact that we made a best choice for $y$. Using this fact, and the equality $c^{y_1}_{i+1} = c^y_i$, it follows that $c^y_{i+2}$ is the unique color which is not in the set $\{c^y_i, c^y_{i+1}, c^z_{i+1}, c^z_{i+2} \}$, or stated differently, the colors $c^y_i, c^y_{i+1}, c^z_{i+1}, c^y_{i+2}, c^z_{i+2}$ are all pairwise disjoint. Writing this property for the index $i+1$ gives that $c^y_{i+1}, c^y_{i+2}, c^z_{i+2}, c^y_{i+3}, c^z_{i+3}$ are all pairwise disjoint. With the overlap between these two sets, it follows that we must have $\{c^y_i, c^z_{i+1}\} = \{c^y_{i+3}, c^z_{i+3}\}$. Moreover, since $c^z_{i+1} \neq c^z_{i+3}$ by assumption on the sequence $\mathcal S'$, then we must have $c^z_{i+3} = c^y_i$, and $c^z_{i+1} = c^y_{i+3}$. Hence, we can see from these conditions that the vertex $y$ takes the colors $1, 2, 3, 4, 5$ successively (up to a permutation of colors), and similarly for $y_1$ and $z$, but with a shift of $1$ and $3$ respectively compared to $y$.

\begin{figure}
    \centering
    \includegraphics{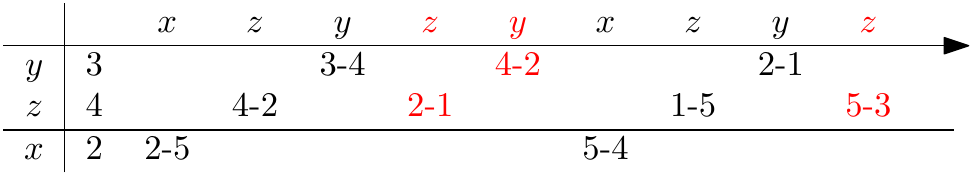}
    \caption{Example for $\mathcal{S'}_{|\{y,z\}}$ and the best choices it implies for $x$. Initial colors of $x$, $y$, and $z$ are $2$, $3$ and $4$ respectively. Recolorings are ordered from left to right. The recolorings in red are saved for $x$.}
    \label{fig:example}
\end{figure}

We can now use the fact that $\mathcal S'$ is very constrained to show that there must be a sufficient number of saved recolorings for $x$, which will contradict the assumption that $x$ is recolored at least $c+1$ times. Let us first assume that there is a recoloring of $x$ caused by the recoloring of $y$ from $c^y_i$ to $c^y_{i+1}$. This means that $x$ is colored $c^y_{i+1}$, and is recolored with a color not in the set $\{c^y_{i}, c^y_{i+1}, c^z_i, c^z_{i+1}\}$. However, we also know that $c^y_{i+2} = c^z_i$, which implies that the recoloring of $y$ from $c^y_{i+1}$ to $c^y_{i+2}$ will be saved for $x$. In a similar way, if the recoloring of $x$ is caused by a recoloring of $z$ from $c^z_i$ to $c^z_{i+1}$, then the new color of $x$ will not be in the set $\{c^z_{i}, c^z_{i+1}, c^y_{i+1}, c^y_{i+2}\}$. And since $c^z_{i} = c^y_{i+2}$, the recoloring of $z$ from $c^z_{i+1}$ to $c^z_{i+2}$ will also be saved. An example of such a sequence $\mathcal{S'}$ is given in Figure \ref{fig:example}.

Hence, in the sequence $\mathcal S'$, for every recoloring of $x$, there is at least one recoloring saved for $x$. Hence, either $x$ is recolored at most $c' - 2$ times during $\mathcal S'$, in which case there are at least $2$ saved recolorings for $x$, or $x$ is recolored at least $c'-2 = 3$ times during $\mathcal S'$, and by the argument above at least two of these recolorings cause a recoloring saved for $x$ (the last one might be at the end of $\mathcal S'$). In all cases, we obtain two saved recolorings for $x$, which is a contradiction of the assumption that $x$ is recolored at least $c+1$ times. Hence $x$ is recolored at most $c$ times and the inductive step holds.

\bibliography{biblio.bib}

\bibliographystyle{abbrv}

\end{document}